\documentstyle[aps,epsf,amsmath]{revtex}

\newcommand{\lp}{l^{+}}
\newcommand{\lm}{l^{-}}
\newcommand{\np}{n^{+}}
\newcommand{\nm}{n^{-}}
\newcommand{\avg}[1]{\langle #1 \rangle}

\begin{document}

\draft \title{Aharonov--Bohm effect in one--channel weakly disordered
  rings} \author{E. P. Nakhmedov\cite{byline}, H. Feldmann, and R.
  Oppermann} \address{Institut f\"ur Theoretische Physik, Univ.
  W\"urzburg,D--97074 W\"urzburg, FRG} \date{}

\maketitle

\begin{abstract}
  A new diagrammatic method, which is a reformulation of Berezinskii's
  technique, is constructed to study the density of electronic states
  $\rho(\epsilon,\phi)$ of a one--channel weakly disordered ring,
  threaded by an external magnetic flux. The exact result obtained for
  the density of states shows an oscillation of $\rho(\epsilon,\phi)$
  with a period of the flux quantum $\phi_0=\frac{hc}{e}$. As the
  sample length (or the impurity concentration) is reduced, a
  transition takes place from the weak localization regime ($L \gg l$)
  to the ballistic one ($L \leq l$). The analytical expression for the
  density of states shows the exact dependence of
  $\rho(\epsilon,\phi)$ on the ring's circumference and on disorder
  strength for both regimes.
\end{abstract}


The oscillation of physical properties of disordered metals has been
studied intensively after the prediction of the Aharonov--Bohm effect
in doubly connected dirty systems \cite{altshuler81a} with the period
of half of a flux quantum and its observation \cite{sharvin81a} in a
Mg cylinder.

Today, a particular subject of intensive investigation is the
persistent current, predicted in \onlinecite{landauer83a,landauer85a}
for one--dimensional disordered rings. Recent advances in
microstructure technology facilitate the fabrication of mesoscopic
rings and the observation of thermodynamic currents therein
\cite{webb91a,mailly93a,levy90a}.  The observed oscillatory responses
in these experiments, which are consistent with a persistent current,
differ in the period of oscillation.

A similar controversy exists also in theory. According to fundamental
physical principles all physical parameters, in particular the
persistent current, of a one--channel metal ring should be periodic in
an applied magnetic flux $\phi$ with period of a flux quantum
$\phi_0=\frac{hc}{e}$
\cite{landauer83a,landauer85a,riedel88a,riedel89a,imry97a}.  However,
the coherent backscattering mechanism with consequent interference
effects in mesoscopic systems gives rise to conductance oscillations
with the halved period $\phi_0/2$
\cite{altshuler81a,montambaux90a,montambaux91a,efetov91a,efetov92a,weidenmueller98a}. It is
pertinent to notice that the attempt to explain the $\phi_0/2$
oscillation in a disordered ring by taking into account the
electron--electron interaction
\cite{weidenmueller98a,eckern90a,schmid91a,riedel91a,kopietz93a} is
also based on the ``cooperon'' propagation in the system.

All these disputes in the theory seem to be connected with the absence
of a consistent theory for a one-dimensional (1d) disordered ring in a
magnetic field which goes beyond the diffusion approximation and can
calculate not only average values of the physical parameters but also
mesoscopic fluctuations of these parameters.

It is well known that the physical parameters of a mesoscopic system
with dimension $L$ satisfying the condition $l < L \ll l_{\rm in}$
(where $l$ is the mean free path and $l_{\rm in}$ is the length over
which the phase coherence of an electron wave is conserved) have
random character, i.e. self--averaging is violated\cite{webb91b}. At
$T=0$ all systems become mesoscopic. In this case high moments give a
considerable contribution, which results in strong differences between
average value and typical one of the observed
parameter\cite{nakhmedov90a}, i.e. the average value loses its
significance to characterize the experimental observation. For such a
problem one has to calculate the whole distribution function and to
get the typical value for an observable
parameter\cite{nakhmedov90a,altshuler89a}.

To perform the procedure presented above there exist technical
difficulties. As far as the Aharonov--Bohm problem for a sufficiently
narrow ring is 1d, the diffusion approach does not give correct
results because of strong interference effects independent of the
degree of randomness \cite{mott61a}. The periodicity adds an
additional technical difficulty.

In this paper we present a new diagrammatic technique by means of
which all diagrams can be summed exactly for weak disorder, when the
criterion $k_F l \gg 1$ (where $k_F$ is the Fermi momentum) is
satisfied.  This method is a generalization of the Berezinskii method
\cite{berezinskii73a}, which was previously developed for a strictly
1d system.

The latter system with $\delta$--correlated Gaussian impurity
potential was studied a long time ago by Halperin \cite{halperin65a}.
In difference to our case, Halperin considered the limit of an
infinite density of scatterers, where the Ioffe--Regel criterion ($k_F
l \approx 1$) is reached. Halperins result describes the energy
dependence of the density of states (DoS) of bound states appearing in
the impurity tail with negative energies.  The same results for the
DoS, together with new information on the localization length,
dielectric constant, and conductivity, were later obtained in an
extension of Berezinskiis theory to strong disorder,
\cite{gogolin82b,gogolin82}.

Here, we consider a one--channel metal ring, threaded by a constant
magnetic flux $\phi$ through the opening. The electrons inside the
ring with circumference $L$ are elastically scattered through the
impurity potential $V_{\rm imp}(x)$.  The Hamiltonian of the system is
written in the form
\begin{equation}
H = \frac{\hbar^2}{2 m^*}\bigl(i\frac{\partial}{\partial x}
+ \frac{2 \pi}{L} \frac{\phi}{\phi_0}\bigr)^2 + V_{\rm imp}(x)
\label{eq:hamiltonian}
\end{equation}
where $x=\varphi \frac{L}{2\pi}$ is the spatial variable on the ring,
$\phi_0=\frac{hc}{e}$ is the fundamental period of a flux quantum and
$m^*$ is the effective mass of an electron. The impurity potential
$V_{\rm imp}(x)$ is considered here to be Gaussian distributed
with a spatial width small enough to justify the Born
approximation.  We apply here our new diagrammatic method to study the
DoS at $T=0$ according to the expression
\begin{equation}
\rho(\epsilon,\phi;x)=-\frac1{\pi} {\rm Im} \avg{G^+(x,x;\epsilon)}
\label{eq:rho}
\end{equation}
where $G^+(x,x';\epsilon)$ is the retarded Green's function (GF) and
the bracket means averaging over the impurity realizations.

Berezinskii's idea to construct a real space diagrammatic method in
one dimension is based on the factorable form of the ``bare'' GF.
However, for a 1d problem with periodic boundary condition, the
quantization of the energy spectrum creates difficulties in this
respect.  To avoid these difficulties, the boundary condition is not
imposed at the beginning and we start with a free particle of
  energy $\epsilon_k = \frac{\hbar^2}{2 m^*}(k - \frac{2\pi}{L}
\frac{\phi}{\phi_0})^2$ and wave function $\Psi_k(x) \propto \exp(i k x)$ with
  continuous $k$.  To implement the periodic boundary conditions,
the particle is allowed to make an arbitrary number of revolutions
around the ring in both directions.  By this means, the ``bare''
retarded Green's function in the coordinate representation can be
expressed in factorable form, as it takes place in the Berezinskii
technique \cite{berezinskii73a}

\begin{equation}
G_0^+ (x,x';\epsilon,\phi)
= \int \frac{dk}{2\pi} \frac{e^{i k (x-x')}}{\epsilon - \epsilon_k \pm i \eta}
=\frac{-i}{\hbar v(\epsilon)}
\exp \left( i 2 \pi \frac{\phi}{\phi_0} 
\frac{x-x'}{L} + i p(\epsilon) |x-x'|
- \frac{\eta}{v(\epsilon)}|x-x'|
\right)
\label{eq:factorable}
\end{equation}

where the dissipative parameter $\eta$ characterizes an energy level
broadening due to inelastic scattering, $v(\epsilon)$ and $\hbar p
(\epsilon)$ are the velocity and the momentum of an electron with
energy $\epsilon$ in a strictly 1d system, respectively, with
$v(\epsilon)=\sqrt{\frac{2 \epsilon}{m^*}}$,
$p(\epsilon)=\sqrt{\frac{2 m^* \epsilon} {\hbar^2}}$.  It is
worthwhile to note that the true GF for a clean ring,
$\tilde{G}_0^+$, can be obtained from Eq.(\ref{eq:factorable}) by
making allowance for arbitrary revolutions.
$\tilde{G}_0^+(x,x';\epsilon,\phi)$ can then be expressed in terms of
$G_0^+$ as
\begin{equation}
\tilde{G}_0^+(x,x';\epsilon,\phi) = \sum_{n=-\infty}^{\infty} G_0^+ (x,x'+n L;\epsilon,\phi)
\label{eq:pureG}
\end{equation}
We can easily verify according to Eqs.(\ref{eq:rho})-(\ref{eq:pureG})
that (the impurity averaging in Eq.(\ref{eq:rho}) of course loses
meaning in this case) the DoS of a clean ring in the presence of an
external magnetic field is
\begin{equation}
\rho_0(\epsilon,\phi)=\rho_0+2\rho_0\sum_{n=1}^{\infty}\cos(p(\epsilon)
L n) \cos(2 \pi \frac{\phi}{\phi_0} n) e^{-\frac{\eta}{v(\epsilon)}Ln}
\label{eq:pureDoS}
\end{equation}
where the DoS of a strictly 1d system is denoted by
$\rho_0=\frac{1}{\pi v(\epsilon) \hbar}$. In the limit $\eta \to 0$,
the DoS assumes the expected discrete form.

\begin{figure}
  \centerline{ \epsfxsize=8cm \epsfbox{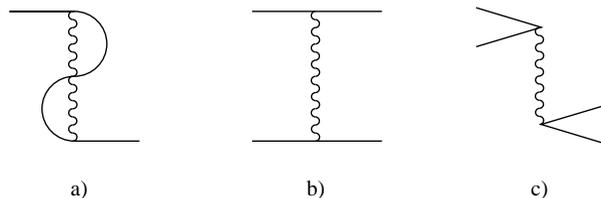}}
\caption{The three internal vertices 
  giving an essential contribution to the impurity averaged GF
in the weak disorder limit $p_{F}l \gg 1$.}
\label{fig:int_vertices}
\end{figure}

Now we represent the unaveraged retarded GF $G^+(x,x';\epsilon,\phi)$ of an electron moving in a
field of randomly distributed impurities by
a continuous line going from point $x$ to $x'$. For the DoS problem,
$G^+(x,x;\epsilon,\phi)$ is adequate.  The factorable structure of the
``bare'' GF $G_0^+ (x,x';\epsilon,\phi)$ between two subsequent
scattering points $x_i$ and $x_{i+1}$ makes it possible to transform
the coordinate dependence from the line to the impurity scattering
points $x_i$, where the impurities are located.  For
our case of weak disorder, the diagrams have to be selected
taking into account $p_{F}l\gg 1$, where $p_{F}$ is the Fermi momentum
and $l$ is the mean free path.  The correlator, connecting two
scattering events and depicted in the diagrams as a wavy
line, characterizes the essential vertices shown in
Fig.\ref{fig:int_vertices}. The expressions corresponding to these
internal vertices are a) $-\left(\frac1{2 \lm}+ \frac1{2 \lp}
\right)$, b) $-\frac1{\lp}$, and c) $-\frac1{\lm}$, respectively,
where $\lp$ and $\lm$ are the mean free paths with respect to forward
and backward scattering \cite{gogolin82}.  The contribution
  of the neglected vertices vanishes, since they oscillate strongly
  with the position\cite{berezinskii73a}.  These expressions show that the internal
vertices don't depend on the magnetic flux $\phi$ and the magnetic
field dependence can be transfered from the internal vertices to the
external ones, which are given in Fig.\ref{fig:ext_vertices}.

\begin{figure}
        \begin{align*}
          \mbox{a)} \setlength{\unitlength}{.6mm}
\begin{picture}(33,10)
\thicklines
\put(8,5){\circle*{4}}
\put(8,5){\line( 1, 0){20}}
\put(23,5){\line( -5,2){7}}
\put(23,5){\line(-5,-2){7}}
\end{picture}
&\sim \sqrt{-\frac{i}{v(\epsilon)}} \exp\bigl(- i 2 \pi
\frac{\phi}{\phi_{0}}\frac{x}{L} + i p(\epsilon)x -
\frac{\eta}{v(\epsilon)}x\bigr)\\
\mbox{b)} \setlength{\unitlength}{.6mm}
\begin{picture}(33,10)
\thicklines
\put(8,5){\circle*{4}}
\put(8,5){\line( 1, 0){20}}
\put(16,5){\line(5,2){7}}
\put(16,5){\line(5,-2){7}}
\end{picture}
&\sim \sqrt{-\frac{i}{v(\epsilon)}} \exp\bigl(+i 2 \pi
\frac{\phi}{\phi_{0}}\frac{x}{L} + i p(\epsilon)x -
\frac{\eta}{v(\epsilon)}x\bigr)\\
\mbox{c)} \setlength{\unitlength}{.6mm}
\begin{picture}(33,10)
\thicklines
\put(28,5){\circle*{4}}
\put(8,5){\line( 1, 0){20}}
\put(21,5){\line( -5,2){7}}
\put(21,5){\line(-5,-2){7}}
\end{picture}
&\sim \sqrt{-\frac{i}{v(\epsilon)}} \exp\bigl(+ i 2 \pi
\frac{\phi}{\phi_{0}}\frac{x}{L} - i p(\epsilon)x +
\frac{\eta}{v(\epsilon)}x\bigr)\\
\mbox{d)} \setlength{\unitlength}{.6mm}
\begin{picture}(33,10)
\thicklines
\put(28,5){\circle*{4}}
\put(8,5){\line( 1, 0){20}}
\put(14,5){\line(5,2){7}}
\put(14,5){\line(5,-2){7}}
\end{picture}
&\sim \sqrt{-\frac{i}{v(\epsilon)}} \exp\bigl(- i 2 \pi
\frac{\phi}{\phi_{0}}\frac{x}{L} - i p(\epsilon)x +
\frac{\eta}{v(\epsilon)}x\bigr)\\
        \end{align*}
\caption{The external outgoing (a,d) and incoming (b,c) vertices and 
  the expressions corresponding to them.}
\label{fig:ext_vertices}
\end{figure}

\begin{figure}
  \centerline{ \epsfxsize=8cm \epsfbox{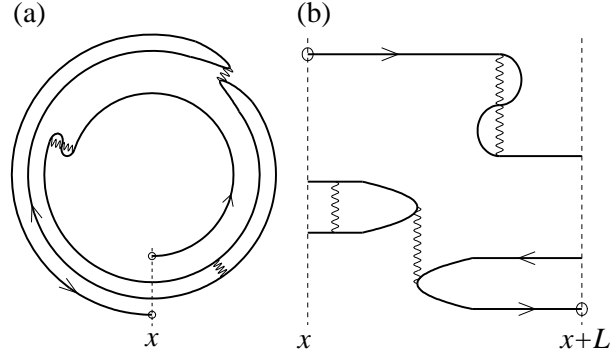}} \vspace{0.3cm}
\caption{(a) A diagram giving a contribution to the DoS. The radial
  unfolding of the drawing was done for the sake of clarity. (b) The
  same diagram as in (a) after cutting at the point $x$.  It belongs
  to the class of diagrams with ($m=1$,$\np=1$,$\nm=0$).}
\label{fig:diagram}
\end{figure}

In difference to two--point correlator problems, the DoS problem can
be described by rather simple diagrams, an example of which is
presented in Fig.\ref{fig:diagram}a. It is convenient to cut the
diagrams at point $x$ and straighten the lines to arrive at the form
shown in Fig.\ref{fig:diagram}b.

Each diagram is characterized by the number of line pairs $m$
returning to the cutting point $x$ and by the total number of
throughgoing lines $n=\np + \nm$, where $\np$ and $\nm$ are the
numbers of rightgoing and leftgoing lines, respectively. According to
Berezinskii's method \cite{berezinskii73a}, the sum of all diagrams
having $m$ pairs of returning lines and $n$ throughgoing lines at the
cross section $x$ is denoted by $Q_{0}(m,n;x-x')$. This block does not
contain the contributions of the external incoming and outgoing
vertices, which are included in the final expression as additional
multipliers.  Due to the structure of the essential vertices,
the number of loops on the left hand side is identical to the number
of loops on the right hand side and we can use the common symbol $m$.

The expression for the average value of the retarded GF can be written
as
\begin{equation}
\begin{split}
  \avg{G^+(\epsilon, \phi;x,x)} = -\frac{i}{v(\epsilon)}
  \sum_{m=0}^\infty \sum_{\np=0}^\infty \sum_{\nm=0}^\infty \Bigl[
  \binom{m + \np}{m} \binom{m - 1 + \nm}{m-1} +\binom{m + \nm}{m}
  \binom{m - 1 + \np}{m-1}
  - \delta_{m,0} \delta_{\np,0}\delta_{\nm,0} \Bigr]\\
  \exp\bigl(i p(\epsilon) L (\np + \nm) -\frac{\eta
    L}{v(\epsilon)}(\np + \nm)- 2 \pi i\frac{\phi}{\phi_0}(\np -
  \nm)\bigr) Q_0(m,n=\np + \nm;L)
\end{split}
\label{eq:GF}
\end{equation}
where the combinatorical factor in the angular brackets
  denotes the different possibilities of ordering the loops and lines:
  assuming that the electron starts from the left hand side, and
  pursuing the continuity of an electron line for the GF, the $\np$
  rightgoing lines can be distributed arbitrarily on the $m+1$
  positions before each of the loops on the left side and directly
  before the final external vertex. Also, the $\nm$ leftgoing lines
  can be distributed on the $m$ positions before the loops on the
  right side.  This gives the first term in the angular brackets of
  Eq.(\ref{eq:GF}) from
\begin{equation}
        \sum_{\{n_{i}^+\}=0}^{\infty} 
        \delta_{\np,n_{1}^++n_{2}^++\dots+n_{m+1}^+}
        \sum_{\{n_{i}^-\}=0}^{\infty} 
        \delta_{\nm,n_{1}^-+n_{2}^-+\dots+n_{m}^-}
        =\binom{m + \np}{m} \binom{m - 1 + \nm}{m-1}
\end{equation}
Similarly, the electron can also start from the right hand side.  In
this case, the left- and rightgoing lines reverse their role, and this
makes the second term in the angular brackets of (\ref{eq:GF}).
However, if there are no loops and no lines, these two cases can not
be distinguished, therefore one has to include the $\delta$--term
in the angular brackets as compensation.  The exponential term in
Eq.(\ref{eq:GF}) comes from the external vertices by taking into
consideration the circulations around the ring.

For the final expression for the DoS, we insert Eq.(\ref{eq:GF}) into
Eq.(\ref{eq:rho}). After some transformation of variables we
obtain

\begin{equation}
\begin{split}
  \rho(\epsilon,\phi) = \rho_0 \sum_{m=0}^\infty \sum_{n=0}^\infty
  \sum_{k=0}^n \Bigl[ 2 \binom{m + k}{m} \binom{m - 1 + n-k}{m-1}
  - \delta_{m,0} \delta_{n,0} \Bigr]\\
  \cos(p(\epsilon) L n) \exp(-\frac{\eta L n}{v(\epsilon)}) \cos(2
  \pi \frac{\phi}{\phi_0}(2k-n)) Q_0(m,n;L)
\end{split}
\label{eq:rho_final}
\end{equation}

\begin{figure}
  \centerline{\epsfxsize=8cm \epsfbox{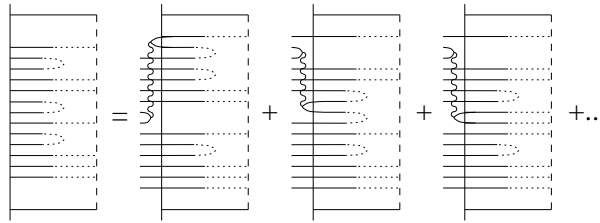}}
\caption{Scheme to construct the equation for the central block $Q_0(m,n;x)$}
\label{fig:insertion}
\end{figure}

The equation for the central block $Q_0(m,n;x)$ is constructed by
infinitesimal shifting the point $x$ and examining the change in $Q_0$
due to passing of the individual impurity lines through $x$
\cite{berezinskii73a,gogolin82}. This process is schematically
presented in Fig.\ref{fig:insertion}. The numbers of possible
insertions of the vertices a) and b) of Fig.\ref{fig:int_vertices} are
$(2m+n)$ and $\frac12 (2m+n)(2m+n-1)$, respectively. The vertex c),
however, can be inserted in two different ways: i) without changing
$m$ and $n$, this can be done in $m(m+n-1)$ different ways (First two
blocks in Fig.\ref{fig:insertion}); and ii) with changing $m$ to $m-1$
and $n$ to $n+2$ (Third block in Fig.\ref{fig:insertion}). The latter
way of insertion has $m^2$ possibilities.

In total, the equation for $Q_0$ is
\begin{equation}
\begin{split}
  &\frac{d}{d x}Q_0(m,n;x) =\\
  & -\Bigl( \frac{(2 m + n)^2}{2 \lp}
  + \frac{n}{2 \lm} + \frac{ m(m+n)}{\lm} \Bigr) Q_0(m,n;x)\\
  & - \frac{1}{\lm} m^2\; Q_0(m-1,n+2;x)
\end{split}
\label{eq:Q}
\end{equation}
$Q_0$ satisfies the boundary condition
\begin{equation}
Q_0(m,n;x=0)=\delta_{m,0}
\label{eq:boundary}
\end{equation}

which means the absence of scattering for a ring with an infinitesimal
small circumference.

To solve Eq.(\ref{eq:Q}), we replace $Q_0(m,n;x)$ according to
\begin{equation}
Q_0(m,n;x)=\exp\bigl(-\frac{x}{2\lm}(2m+n)^2 - \frac{x}{\lm}m(m+n) - \frac{x}{2\lm}n\bigr)\tilde{Q}_0(m,n;x)
\label{eq:Q_substitution}
\end{equation}

By Laplace transforming $\tilde{Q}_0$ from the coordinate $x$ to
  the new variable $\lambda$ and by using the boundary condition
(\ref{eq:boundary}), the equation for $Q_0$ is reduced to the form
\begin{equation}
\lambda \overline{Q}_0(m,n;\lambda)-\delta_{m,0}=
-\frac{1}{\lm} m^2 \exp\bigl(\frac{xn}{\lm}\bigr)
\overline{Q}_0(m-1,n+2;\lambda-\frac{n}{\lm})
\label{eq:laplace}
\end{equation}
Eq.(\ref{eq:laplace}) can be solved by iteration in $m$. For $m=0$,
$\overline{Q}_0(0,n;\lambda)=\frac{1}{\lambda}$. Further iteration
gives
\begin{equation}
\overline{Q}_0(m,n;\lambda)=\frac{(-1)^m(m!)^2}{(\lm)^m}\prod_{j=0}^{m}\frac{1}{\lambda-\frac{1}{\lm}j(j+n-1)}
\end{equation}
Inverse Laplace transform of $\overline{Q}_0(m,n;\lambda)$ results in
\begin{equation}
\begin{split}
  Q_0(m,n;L)=\exp\bigl(-\frac{L}{2\lp}(2m+n)^2 - \frac{L}{\lm}m(m+n)-\frac{L}{2\lm}n\bigr)\\
  \sum_{j=0}^m(-1)^j\binom{m}{j}\frac{m!(j+n-2)!}{(m+j+n-1)!}(2j+n-1)
  \exp\bigl(\frac{L}{\lm}j(j+n-1)\bigr)
\end{split}
\label{eq:Q_sol}
\end{equation}
where the exponential prefactor in (\ref{eq:Q_substitution}) has been
taken into account.
From Eq.(\ref{eq:Q_sol}) it can be verified that for $L=0$ the sum over $j$ gives $Q_0(m,n,L=0)=\delta_{m,0}$ and that $Q_0$ decays exponentially with $n$ for $m=0$. Also, one obtains from (\ref{eq:Q_sol}) in the special case of $n=0$
\begin{equation}
Q_0(m,n=0;L)=(1-m-m\frac{L}{\lm})e^{-\frac{2L}{\lp} m^2 - \frac{L}{\lm}m^2}+e^{-\frac{2L}{\lp}m^2-\frac{L}{\lm}m^2-\frac{L}{4\lm}}\sum_{j=2}^m (-1)^j \binom{m}{j}\frac{m!(j-2)!}{(m+j-1)!}(2j-1)e^{\frac{L}{4\lm}(2j-1)^2}
\end{equation}

Eqs.(\ref{eq:rho_final}) and (\ref{eq:Q_sol}) constitute the exact
result for the DoS of a one--channel weakly disordered ring in an
external magnetic field. The result is valid for weak localization
and ballistic regimes.

\begin{figure}
  \centerline{\epsfxsize=10cm \epsfbox[240 449 582 669]{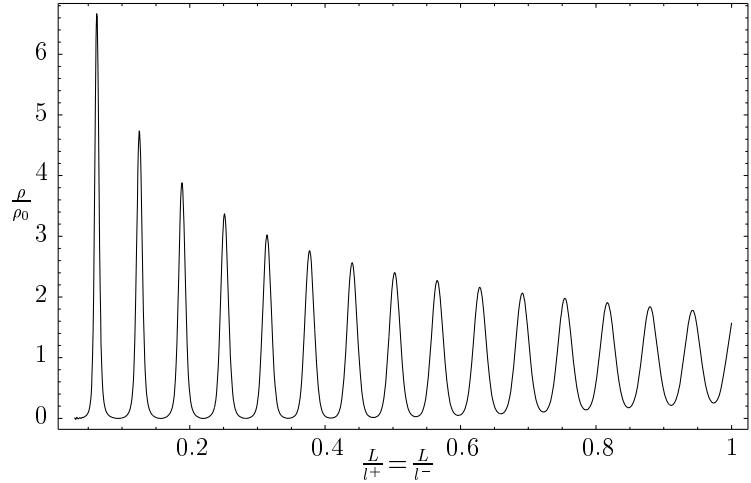}}
\caption{Dependence of the DOS on the ring length $L$ for small rings and zero magnetic field, obtained from Eqs.(\ref{eq:rho_final}) and (\ref{eq:Q_sol}). The fermi momentum is given by $p l^- = 100$.
The sharp discrete levels [Eq.(\ref{eq:rhoBallistic})] for small ring length or large scattering length cross over to a continuous DOS for large rings [Eq.(\ref{eq:rho_weak_localization})].}
\label{fig:dos_small_rings}
\end{figure}

For the weak localization regime, corresponding to the criterion
$L \gg \max \{\lp,\lm\}$,
Eqs.(\ref{eq:rho_final}) and (\ref{eq:Q_sol}) are simplified to
\begin{equation}
\begin{split}
  \rho(&\epsilon, \phi)=\rho_0 \Bigl\{1-\frac{2L}{\lm} \exp
  \bigl(-\frac{2L}{\lp}-\frac{L}{\lm} - \frac{\eta}{v(\epsilon)}L\bigr)\Bigr\} \\
  + & 2 \rho_0 \exp\bigl(-\frac{L}{2 \lp}-\frac{L}{2 \lm} - \frac{\eta}{v(\epsilon)}L\bigr) \Bigl
  \{
  \cos(p(\epsilon) L) \cos(2 \pi \frac{\phi}{\phi_0})
   + \exp \bigl(-\frac{3 L}{2 \lp}-\frac{L}{2 \lm} \bigr) \cos(2
  p(\epsilon) L) \cos(4 \pi \frac{\phi}{\phi_0}) \Bigr \}
\end{split}
\label{eq:rho_weak_localization}
\end{equation}
which shows that the leading contribution to the DoS oscillation has a
period of $\phi_0$ and its amplitude decreases
exponentially with impurity strength (or with increasing $L$) for weak
disorder.  Such a small contribution of the impurity scattering to the
DoS is connected with the absence of ``diffusion'' and ``cooperon''
contributions to the averaged Green's functions. In the limit of an
infinite sample, the correction due to weak disorder disappears
completely. For the ballistic regime, when $L \leq \min \{\lp,\lm \}$,
the contribution to the DoS can be approximated in the form
\begin{equation}
\begin{split}
  &\rho(\epsilon,\phi)=\rho_0(\epsilon,\phi)-\rho_0 \frac{L}{\lp}
  \sum_{n=0}^{N_+}
  n^2 \cos(p(\epsilon) L n) \cos(2 \pi \frac{\phi}{\phi_0} n)
e^{-\frac{\eta}{v(\epsilon)}Ln}\\
  &- \rho_0 \frac{L}{\lm} \sum_{n=0}^{N_-} n \cos(p(\epsilon) L n)
  \cos(2 \pi \frac{\phi}{\phi_0} n)e^{-\frac{\eta}{v(\epsilon)}Ln}\\
  & - 2 \rho_0 \frac{L}{\lm} \sum_{n=0}^{N_-} \sum_{k=0}^{k=n} (k+1)
  \cos(p(\epsilon) L n) \cos(2 \pi \frac{\phi}{\phi_0}(2k-n))e^{-\frac{\eta}{v(\epsilon)}Ln}
\end{split}
\label{eq:rhoBallistic}
\end{equation}
where $\rho_0(\epsilon,\phi)$ is the DoS of a clean ring given by Eq.(\ref{eq:pureDoS}), $N_+\approx \bigl[ \sqrt{\frac{2 \lp}{L}} \bigr]$ and
$N_-\approx \bigl[ \frac{2 \lm}{L} \bigr]$.  From
Eq.(\ref{eq:rhoBallistic}) it can be seen that disorder gives a
contribution proportional to $\frac{L}{l^\pm}$ to $\rho_0(\epsilon,\phi)$.
the DoS of a clean
system in a magnetic field,
$\rho_0(\epsilon,\phi)=\rho_0+2\rho_0\sum_{n=1}^{\infty}\cos(p(\epsilon)
L n) \cos(2 \pi \frac{\phi}{\phi_0} n)$. The dependence of the DOS on the ring circumference $L$, computed from Eqs.(\ref{eq:rho_final}) and (\ref{eq:Q_sol}), is given in Fig.\ref{fig:dos_small_rings}, where the 
sharp discrete peaks in the ballistic regime are due to energy level quantization.

The upper limit $N_\pm$ of
the sums in Eq.(\ref{eq:rhoBallistic}) may be a small value, e.g.
$N_\pm \approx 1$, deduced from $\frac{l}{L}=1.3$ according to the
experiment in Ref. \onlinecite{mailly93a}. Therefore, the oscillation
with a full flux quantum $\phi_0$ will be pronounced in the ballistic
regime.

In the absence of backward scattering ($\lm = \infty$) in the system,
Eqs.(\ref{eq:rho_final}) and (\ref{eq:Q_sol}) give a rather simple
expression for the DoS, which can be presented in the following form:
\begin{equation}
\begin{split}
  \rho(\epsilon,\phi)&=\rho_0+\frac{\rho_0}{2} \sqrt{\frac{\lp}{2 \pi
      L}}
  \int_{-\infty}^{\infty} d\gamma e^{-\frac{\lp}{2 L} \gamma^2}\\
  & \Bigl(\frac{1}{\exp[-ip(\epsilon)L - i 2 \pi \phi/\phi_0 +i \gamma + \frac{\eta}{v(\epsilon)}L]-1}+\frac{1}{\exp[-ip(\epsilon)L + i 2 \pi \phi/\phi_0 +i \gamma + \frac{\eta}{v(\epsilon)}L]-1}
+ \mbox{c.c.}\Bigr)
\end{split}
\label{eq:onlyForward}
\end{equation}
Eq.(\ref{eq:onlyForward}) can be physically interpreted as follows:
each act of forward scattering gives rise to coherent shifting of all
energy levels. The value of this shifting is random with Gaussian
distributions; the typical value of this shifting is proportional to
$\frac{\hbar}{\tau^+}\sqrt{\frac{\lp}{L}}$ where $\tau^+$ is the
relaxation time due to forward scattering. Therefore, only backward
scattering seems to be responsible for level repulsion in disordered
1d systems\cite{altshuler86a}.  Averaging over this random shifting
results in Gaussian broadening of the energy levels, which for the
weak localization regime is much smaller than Dingle broadening and
comparable with it in the ballistic regime. In the latter case, the
transport seems to be connected with resonant tunneling.

\begin{figure}
  \centerline{
\epsfxsize=8cm \epsfbox[294 521 530 669]{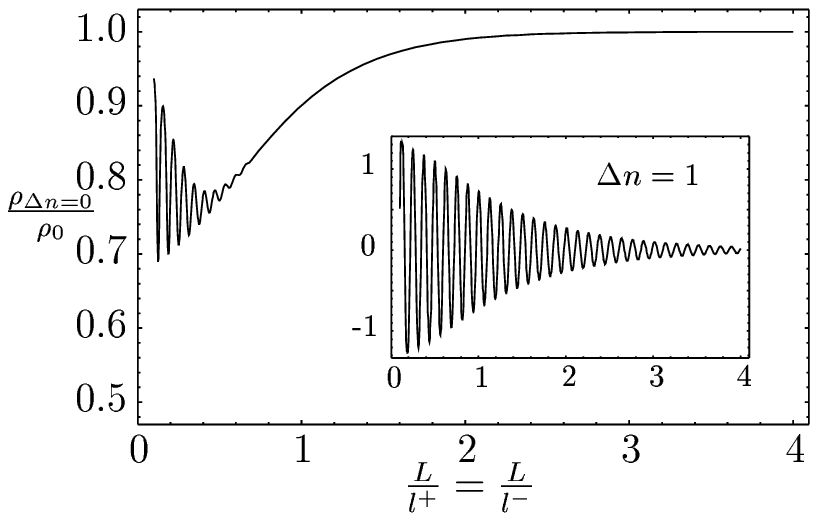}
\epsfxsize=8cm \epsfbox[293 520 530 670]{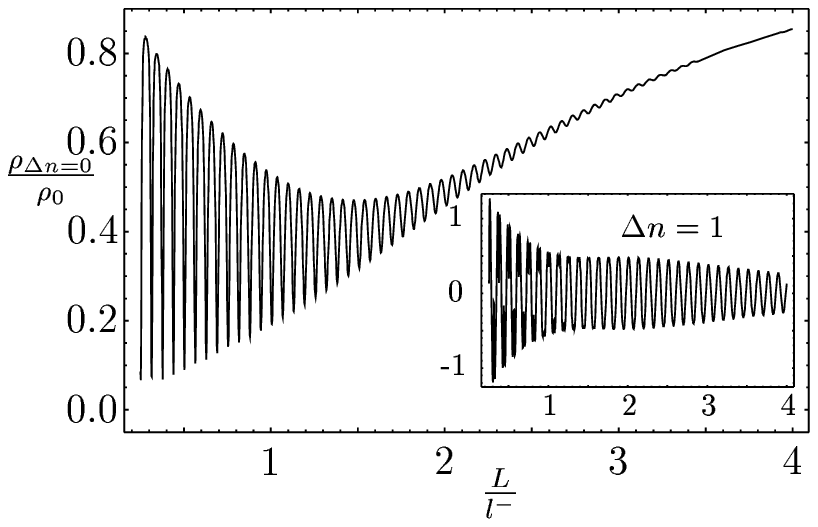}}
\caption{The field independent ($\Delta n=0$) part of the DOS for $\lp=\lm$ (l.h.s) and for $\lp=\infty$ (r.h.s); and $p \lm =50$. The insets show the $\Delta n=1$ contributions. Higher oscillations ($\Delta n > 1$) are similar, but with increased damping.}
\label{fig:dos_orders}
\end{figure}

To illustrate the dependence of $\rho$ on $l^+$ and $l^-$, we decomposed
the DOS [Eq.(\ref{eq:rho_final})] into the field independent part with
the restriction $\Delta n=|\np-\nm|=0$, and the harmonics with $\Delta n=1,2\dots$. For $\lp=\lm$ and for $\lp=\infty$ we show these contributions in Fig.\ref{fig:dos_orders} as functions of the ring length. For $\lm=\infty$, the field independent part is constant and the higher harmonics are simple damped oscillations.

It is necessary to notice that the oscillative behavior of a
persistent current will differ from that obtained for the DoS. In
contrary to the dynamical approach to the study of conductance, as a
result of which the latter is connected with a current--current
correlator, the average thermodynamic current $ \avg{I(\phi)}$ is
defined by the average value of the thermodynamic potential $F$
according to $\avg{I(\phi)} = -c \frac{\partial \avg{F}}{\partial
  \phi}$ where the bracket denotes an averaging over the impurity
realizations.

At zero temperature the last expression turns to
$\avg{I(\phi)}=-c\frac{\partial \avg{E}}{\partial \phi}$ with
$E=\int_0^{\mu(\phi)} d \epsilon\, \epsilon \rho(\epsilon,\phi)$ being
the total energy of the particles. Expressing DoS
$\rho(\epsilon,\phi)$ and the flux dependent Fermi energy $\mu(\phi)$
as $\rho = \avg{\rho} + \delta \rho$ and $\mu=\avg{\mu}+\delta\mu$
where $\avg{\delta \rho}=\avg{\delta\mu}=0$ and using in addition the
particle number conservation $N={\rm const}=\int_0^{\mu(\phi)}
d\epsilon\,\rho(\epsilon,\phi)$ to determine $\delta \mu$, $\avg{E}$
can be written in the following form:
\begin{equation}
\begin{split}
  \avg{E}=&\int_0^{\mu_0}d\epsilon\,\epsilon\avg{\rho(\epsilon,\phi)}-
  \frac{\mu_0}{\avg{\rho}}
  \int_0^{\mu_0}d\epsilon\,\avg{\delta\rho(\epsilon,\phi)\delta\rho(\mu_0,\phi)}\\
  &+\frac1{2\avg{\rho}}\Bigl[1+\frac{\mu_0}{\avg{\rho}}
  \frac{\partial\avg{\rho(\mu_0,\phi)}}{\partial\mu_0}\Bigr]
  \int_0^{\mu_0}d\epsilon_1\int_0^{\mu_0}d\epsilon_2\,
  \avg{\delta\rho(\epsilon_1,\phi)\delta\rho(\epsilon_2,\phi)}
\end{split}
\end{equation}
As it is seen from this expression, contributions to the persistent
current are given not only by the average value of the DoS but also by
the correlator $\avg{\delta\rho\delta\rho}$.  By expressing the DoS as
a difference of the retarded ($G^+$) and advanced ($G^-$) Green's
functions, the latter correlator is shown to be dominated by
$\avg{G^+G^-}$. Characterizing the retarded (advanced) Green's
function by the number of line pairs $m$ ($\overline{m}$) and by the
total number of throughgoing lines $n=n^++n^-$
($\overline{n}=\overline{n}^++\overline{n}^-$), the oscillating
factors in the kernel for the correlator $\avg{G^+G^-}$ will have the
form (compare with Eq.(\ref{eq:rho_final}))
$\cos(p(\epsilon)L(n-\overline{n})) \cos(2 \pi
\frac{\phi}{\phi_0}(n^+-n^-+\overline{n}^+-\overline{n}^-))$.  The
main contribution to the correlator comes from the terms with
$n=\overline{n}$, which oscillate with the halved period,
\cite{ComingPaper}.

The explanation presented above for the $\frac{\phi_0}{2}$ --
periodicity of $\avg{I(\phi)}$ is connected with the transition from
the grand canonical ensemble averaging to the canonical one
\cite{riedel88a,montambaux90a,montambaux91a,schmid91a,riedel91a,altshuler91a}. Another consideration also
seems to be pertinent. For a mesoscopic system the total energy $E$ at
$T=0$ is a random parameter. For a fixed chemical potential in the
system the number of levels fluctuates from sample to sample. An
external magnetic field will periodically change these fluctuations.
The correct approach to the problem is to find the distribution
function of $E$, which seems to be the same as the one for the DoS,
and to calculate the typical value of $E$ by means of averaging. Such
averaging will include not only $\avg{\rho}$ but also higher moment
correlators.

The method presented in this paper makes it possible to calculate all these
moments of the DoS \cite{ComingPaper}, as it was done in
\cite{altshuler89a} for a strictly 1d system.
Further, the extention to correlators of the local DOS with 
different energies and different positions allows to study level repulsion.

In conclusion, we have presented a new diagrammatic method which gives
an exact result for a one--channel weakly disordered ring threaded by
a magnetic flux $\phi$. As an application of the method the DoS of an
Aharonov--Bohm ring in the absence of electron--electron interaction
is calculated. The result obtained gives an exact dependence of the
DoS on the parameters $\phi$, $\epsilon$, $L$ and $l^\pm$.

We are thankful for useful discussions with V. N. Prigodin in the
early stage of this work. This work was supported by the SFB410.

% for submission:

% for use of bibtex:
%\bibliographystyle{prsty}
%\bibliography{oppgroup}

\end{document}